# Acoustic Graphene Plasmon Nanoresonators for Field Enhanced Infrared Molecular Spectroscopy

Shu Chen[1,2], Marta Autore[1], Jian Li[1,3], Peining Li[1], Pablo Alonso-Gonzalez[1,7], Zhilin Yang[2], Luis Martin-Moreno[4], Rainer Hillenbrand[1,5,6*] and Alexey Y. Nikitin[1,5*]

[1]CIC nanoGUNE, Donostia-San Sebastián 20018, Spain

[2]Department of Physics, Xiamen University, Xiamen 361005, China

[3]State Key Lab of Analytical Chemistry for Life Science, School of Chemistry and Chemical Engineering, Nanjing University, Nanjing 210093, China

[4]Instituto de Ciencia de Materiales de Aragón and Departamento de Fisica de la Materia Condensada, CSIC-Universidad de Zaragoza, E-50009, Zaragoza, Spain

[5]IKERBASQUE, Basque Foundation for Science, Bilbao 48013, Spain

[6]CIC nanoGUNE and UPV/EHU, Donostia-San Sebastián 20018, Spain

[7]Departamento de Física, Universidad de Oviedo, 33007, Oviedo, Spain

**ABSTRACT**: Field-enhanced infrared molecular spectroscopy has been widely applied in chemical analysis, environment monitoring, and food and drug safety. The sensitivity of molecular spectroscopy critically depends on the electromagnetic field confinement and enhancement in the sensing elements. Here we propose a concept for sensing, consisting of a graphene plasmonic nanoresonator separated from a metallic film by a nanometric spacer. Such a resonator can support acoustic graphene plasmons (AGPs) that provide ultra-confined electromagnetic fields and strong field enhancement. Compared to conventional plasmons in graphene, AGPs exhibit a much higher spontaneous emission rate (reaching values up to $1\times10^8$), higher sensitivity to the dielectric permittivity inside the AGP nanoresonator (figure of merit is higher by a factor of 7) and remarkable capability to enhance molecular vibrational fingerprints, of nanoscale analyte samples. Our work opens novel avenues for sensing of ultra-small volume of molecules, as well as for studying enhanced light-matter interaction, *e.g.* strong coupling applications.

Infrared (IR) molecular spectroscopy plays an important role for variety of applications, for example, chemical analysis, environment monitoring and food safety, since it can provide spectral fingerprint information of molecules in a nondestructive label-free detected way.[1,2] However, the detection of molecules at the nanoscale still poses difficulties, due to the extremely weak vibrational absorption signals originating from the large wavelength of the IR light (μm-scale).[3,4] The efficiency of the interaction of light with molecules is also dramatically dependent upon the volume of the probed material, being poor in case of subwavelength volumes. Strong electromagnetic field enhancement and its confinement at locations of adsorbed molecules are of great importance to overcome the above limitations. Plasmons – collective oscillations of charge carriers coupled to electromagnetic field – can provide both enhancement and confinement of the electromagnetic field, thus offering great opportunities to enhance the interaction between incident light and molecules.[3-5] Plasmons in metals have already been shown to have a strong potential for field enhanced IR molecular spectroscopy.[6-8] However, the weak confinement of plasmons in the mid-IR frequency range challenges the sensing of thin molecular layers. Indeed, to reach a high concentration of the electromagnetic field, one has to design metallic structure with narrow gaps, which is highly demanding from the point of view of fabrication. Another disadvantage of plasmons in metals is their limited tunability (the shaping of the geometry of metallic structures is needed for that).[2,3] Therefore, to create efficient plasmonic sensors in the mid-IR frequency range, a quest for novel materials that can support highly confined and tunable plasmons arises.

Graphene supports plasmons at mid-IR frequencies and has shown a strong potential for many interesting applications,[9,10] such as tunable metamaterials,[11] thermoelectric devices,[12] optical modulators,[13] among others. Graphene plasmons (GPs) are strongly confined and thus can be beneficial for enhancing IR molecular signals, particularly for probing small quantities of molecules.[2, 3,14-17] Moreover, a strong dependence of GPs wavelength upon the voltage applied to graphene -*via* electrostatic gating - allows for a high tunability of potential sensing devices based on GPs. A particularly interesting GP mode – screened graphene plasmons (SGP) – has been predicted theoretically[18-20] and recently demonstrated experimentally in a graphene sheet placed above a metallic substrate (for example, acting as a gate) separated from graphene by a distance $d$.[21] SGPs result from the hybridization of plasmons in the graphene sheet and their mirror image and exhibit linear dispersion.[21] Both mathematically and physically SGPs are equivalent to the acoustic graphene plasmons (AGPs) in the double graphene layer, where the separation between the graphene sheets is $2d$. Due to this equivalence, we will further refer to SGP as AGP. The electromagnetic field of AGPs is squeezed between the graphene and metallic substrate so that the vertical confinement of AGPs is determined essentially by the width of the gap (analogously to plasmons in nm-scale metallic gaps[22]), thus surpassing the diffraction limit by orders of magnitude. Because of their strong confinement, AGPs have several advantages compared to conventional plasmons in graphene without metallic gate, especially for applications requiring strong light-matter interactions. Additionally, the electromagnetic field enhancement and confinement can be further improved by shaping gated graphene into nanoresonantors (for example, disks, rectangles or ribbons) analogously to nanostructured graphene on a dielectric substrate.[23-25]

Here, we study plasmon resonances in graphene disks separated from a metallic substrate by a nanometric spacer. By means of full-wave electromagnetic simulations, we identify a rich variety of AGP modes supported by the nanoresonators. We reveal strong AGP resonance shifts when the dielectric permittivity of the spacer varies and further demonstrate the potential of the AGPs for sensing applications considering a protein molecule as an example.

**RESULTS AND DISSCUSSION**

**Dispersion of AGPs**

Let us start with the comparison of the dispersion and the field distribution of plasmons in graphene on a dielectric substrate (upper schematic in Fig. 1a) – conventional GPs – and graphene separated by a dielectric spacer from a metallic substrate (bottom schematic in Fig. 1a) – AGPs.

Fig. 1c shows the dispersion curves of both AGPs (red and blue curves) and conventional GPs (black curve). One can see that at a fixed frequency, ω, the momentum of AGPs, $k_{AGP}$, is significantly larger than the momentum of conventional GPs, $k_{GP}$. This is also confirmed by the field snapshots (Fig. 1b), where the AGP wavelength, $\lambda_{AGP} = 2\pi/k_{AGP}$ is significantly smaller than that of GPs, $\lambda_{GP} = 2\pi/k_{GP}$. The field of the AGPs is squeezed into the spacer, having capacitor-like distribution. Importantly, the momentum of AGPs increases with the decrease of the thickness d of the spacer, as seen from the comparison of the red and blue curves in Fig. 1c. For small d, the AGP momentum scales as $k_{GP} \propto 1/\sqrt{d}$.[18, 19, 21] Thus, by reducing d it could be possible to confine AGPs to sub-nm gaps between graphene and metal, leading to substantially increase the AGP confinement. The dispersion of AGPs in this case will be governed by quantum non-local effects.[26] Assuming that for a nanometric spacer, the nonlocal effects will not qualitatively change the AGP dispersion, we will neglect the nonlocality in all our simulations.

**Plasmonic modes of AGP disk nanoresonators revealed by the energy loss of a point emitter**

In the following we will study AGP nanoresonators consisting of graphene disk separated by a few-nm dielectric spacer from a metallic substrate. Since AGP modes have large wavevectors, they do not couple to light propagating in free space. Therefore, for studying plasmonic modes in AGP nanoresonators, we will use a vertical point dipole emitter. The latter can generate deeply subwavelength scale electromagnetic fields of both longitudinal and transversal polarizations, and thus can efficiently excite plasmonic modes with arbitrarily large wavevectors and arbitrary symmetry.[23, 24, 27, 28] We start with the characterization of AGP disk nanoresonantor by calculating the energy loss of a vertical electric point dipole source (polarized along the z-axis) placed above the disk (schematic in Fig. 2a). The energy loss of the dipole as a function of frequency will reveal the spectral positions of AGP resonances and help to further classify the corresponding AGP modes. The energy loss can also serve as a measure of the decay rate of a quantum emitter interacting with AGP modes.[23,29]

A doped graphene nanodisk with a diameter $D = 50$ nm is placed 2 nm above a gold substrate with a transparent spacer ($\varepsilon = 1$) in between. Throughout this work, we consider the Fermi energy of the charge carriers in graphene to be $E_F = 0.5$ eV and their relaxation time $\tau = 0.05$ ps, corresponding to a typical CVD graphene (with mobility $\mu = 1000$ cm$^2$/(V·s)).[30-32] The dipole **p** is located at different lateral positions above the disk (marked by red, purple and black dots corresponding to A, B, C in Fig. 2a): at the center (point A), between the center and edge (point B), and the edge of the disk (point C). Its vertical distance to the graphene is 5 nm, which has been chosen to provide an optimal coupling to the AGPs (see the detailed discussion later).[33]

The normalized total decay rate γ can be calculated according to the following equation[27]

$$\gamma = \frac{\Gamma}{\Gamma_0} = -\frac{3c^3 \cdot \text{Re}(E_z)}{2\omega^2} \qquad (1)$$

Where Γ and $\Gamma_0$ are the total energy losses of the dipole above the disk and in free space, respectively. $E_z$ is the z-component of electric field (resulting from the electromagnetic interaction between the dipole and graphene disk) at the position of the dipole, and can be found from Maxwell's equations with the help of full wave simulations (see Methods). In Fig. 2b, we show the calculated decay rate γ as a function of frequency for the three dipole positions A, B and C. We observe that $\gamma$ exhibits several resonant peaks, which we attribute to the excitation of plasmon modes. The number of peaks and their resonant frequencies are different for each dipole position. We label the peaks according to the position of the dipole, by a capital with an integer subscript (i.e. $A_1$, $A_2$ etc.), and proceed with their identification.

To identify the peaks and the corresponding plasmonic modes, we plot the electric field distributions, $Re(E_z)$, in both xy- and xz-plane (Fig. 2c) at the frequency positions of the peaks (the field distributions in Fig. 2c have the same labels as the corresponding peaks in Fig. 2b). Based on the field distributions in the xy-plane, we can easily distinguish two different types of plasmonic modes, analogously to graphene nanoresonators on a dielectric substrate.[24] The modes of the first type (see snapshots labeled as C in Fig. 2c and their corresponding spectral positions in the black curve of Fig. 2b) have the electromagnetic field localized along the edge of the disk, with $Re(E_z)$ showing oscillations that evolve along the disk edge, which lets us to identify them as *edge plasmons*.[24,34] According to the field distribution, the edge plasmons lead to the first-order ($C_1$), second-order ($C_2$), third-order ($C_3$), fourth-order ($C_4$) and fifth-order ($C_5$) whispering-gallery resonances,[24,34,35] respectively. The whispering-gallery edge plasmons resonances are also excited when the dipole is placed at position B. In this case the peaks $B_1$-$B_3$ of the purple spectrum are reminiscent to the peaks $C_1$-$C_3$ of the black spectrum in Fig. 2b. The close frequency positions of the peaks for $B_1$-$B_3$ and $C_1$-$C_3$, as well as the similar field distributions (not shown) indicates that the same modes can be excited for different dipole positions. The modes of the second type (snapshots labeled as $B_4$ and $A_1$ in Fig. 2c), have the electric field oscillating over the whole area of the disks instead of being confined to the edge. Such a distribution of the electric field reveals *sheet modes*.[24,34] Particularly, when the dipole is located at the point A, the field of the sheet mode labeled as $A_1$, at the frequency of 2292 cm$^{-1}$ possesses a circular symmetry, with the field oscillation in the radial direction. Therefore, $A_1$ can be classified as a radial *breathing mode*.[36]

The spectral overlap of the peaks corresponding to different field patterns (e.g. $C_3$ and $A_1$ in the black and red curves in Fig. 2b, respectively) clearly indicate the *coexistence* of different modes, implying a multimode character of plasmonic excitations in the disk. Note that in heterogeneous samples, the coexistence of modes can lead to broadening of resonance for certain dipole locations. Therefore, the high quality of the samples is preferable for a selective coupling to the individual modes. Importantly, according to the field distribution in the xz-plane (Fig. 2c), the near field of all the plasmonic modes is highly concentrated in the dielectric spacer between the graphene and the metallic substrate. Moreover, the electric field inside the spacer shows a $\pi$-phase shift compared to that outside of the spacer, as confirmed by the change of the polarity of $Re(E_z)$ (change of the color). Both the high field concentration inside the spacer and the out-of-phase field oscillations in the xz-plane is consistent with the field distributed feature of AGPs (shown in Fig. 1b), and thus confirm that all the excited plasmonic modes are AGP modes.[21]

For comparison, we also show simulations for a free-standing graphene disk (D = 110 nm) in Fig. 2d-f, supporting GP modes. Fig. 2d shows $\gamma$ as a function of frequency for the dipole located above the disk center (point A). The peaks $A_1'$ and $A_2'$ correspond to the excitation of the first and second-order breathing modes, respectively. Their $Re(E_z)$ distributions are shown in Fig. 2f. Comparing the $Re(E_z)$ distributions in the xz-plane for the modes $A_1$ (Fig. 2c) and $A_1'$ (Fig. 2f), we see that AGP nanoresonators have a stronger electric field confinement in the vertical direction compared to GP nanoresonators. Additionally, the diameter of the AGP disk nanoresonator is smaller by a factor of 2 compared to the GP disk nanoresonator (50 nm vs 110 nm), although in both nanoresonators the first-order breathing mode has the same resonance frequency ($\omega = 2292 \text{ cm}^{-1}$). Such a difference in resonator size is consistent with a similar difference between the wavevectors for AGPs and GPs according to Fig. 1c. Indeed, at the resonance frequency (horizontal dashed line), the ratio between the wavevectors (the wavevectors are indicated by the vertical dashed lines) is $k_{AGP}/k_{GP} = 2.3$. Interestingly, the position of the AGP resonance can be shifted to lower frequencies by decreasing the distance between the graphene disk and the metallic substrate while keeping disk diameter. In contrast, for GP disk nanoresonators, the resonance frequency can be reduced only by increasing the size of disk. Thus, the difference in size between AGP and GP nanoresonators becomes more notable at lower frequencies. The smaller size of AGP nanoresonators (and thus smaller volumes of supported AGP modes) as well as their additional tunability via the spacer thickness can be advantageous for sensing of smaller quantities of molecules.

The amplitude of the decay rate $\gamma$ of the dipole placed above the AGP nanoresonator ($1.8 \times 10^6$ in the maxima of the peak $A_1$ in Fig. 2b) is higher by approximately a factor of two compared to that of GP nanoresonator ($8.8 \times 10^5$ in the maximum of the peak $A_1^{'}$ in Fig. 2e). The higher value of $\gamma$ for the AGP nanoresonator is consistent with the higher field confinement compared to the GP nanoresonator. Note that $\gamma$ strongly depends on the distance from the dipole to the graphene. Fig. 3 shows $\gamma$ as a function of frequency (within 500-3000 cm$^{-1}$ range) for several dipole positions along the z axis (labeled as $z_1$, $z_2$, $z_3$, $z_4$ in Fig. 3a) for the AGP nanoresonator. The selected frequency range and the dipole location (the center of the disk) corresponds to the excitation of the breathing mode $A_1$. As follows from Fig. 3a, $\gamma$ dramatically increases with the decrease of the dipole-graphene distance. In order to clarify the contribution of AGPs to $\gamma$, we show the latter as a function of the distance between the dipole and graphene surface at a fixed frequency $\omega = 2292 \text{ cm}^{-1}$ in Fig. 3b, where a double logarithmic scale is used for a better representation. In the function $\gamma(z)$ we can clearly distinguish 3 regions, where the energy loss is dominated by different processes (see details in Ref. 27[27]): (i) for distances $z > 100$ nm, $\gamma$ is dominated by radiation of propagating waves and rather constant (the green curve coincides with the blue curve in Fig. 3b); (ii) distances z ranging from 1 to 100 nm, where the dipole couples to AGPs, while $\gamma_{GPs} \sim e^{-2k_p \cdot z}$ with $k_p$ being the AGPs wavevector (the green curve coincides with the magenta curve in Fig. 3b);[33] (iii) small distances $z < 1$ nm, where the energy is lost via generation of evanescent fields with high momenta and $\gamma$ diverges in the local approximation (growth of the green curve towards small z in Fig. 3b). Note that the positions of the crossovers between the regions (i)-(ii) and (ii)-(iii) on the green curve are dictated by the vertical confinement of the plasmonic mode, being frequency-dependent. The Fig. 3b thus proves the major role of AGPs in the decay rate for the dipole-disk separations in the range between 100 nm and a few nanometers. Most importantly, the highest values of $\gamma$ are found for the dipole located *inside* the spacer. This is corroborated by the black curve in Fig. 3c (calculated for $z_1 = -1$ nm, corresponding to the dipole below the graphene disk), showing the values of the normalized total decay rate up to

$1.0\times10^8$. Comparing the black with red curves in Fig. 3c, one can conclude that $\gamma$ of the dipole inside the spacer is one order of magnitude higher than that of the dipole located at the same distance ($z_1 = 1$ nm) above the disk. This reveals a much better coupling of AGP modes with the dipole in the region of the spacer, i.e. where the AGP field has the strongest confinement.

**Refractive index sensing with AGP nanoresonators and figure of merit**

The ultra-high field confinement inside the dielectric spacer of AGP disk nanoresonator can be used for refractive index sensing of a small amount of analyte, placed inside of the spacer. To demonstrate the sensing capability of the AGP nanoresonator, in the simulations we monitor the shift of the AGP resonance as a function of the dielectric permittivity of the spacer, $\varepsilon$.[37] Figure 4b illustrates the spectral shift of the AGP resonance (corresponding to $A_1$ mode in Fig. 2b) upon $\varepsilon$ for a 50 nm diameter graphene disk separated by a 2 nm-thick spacer from the metallic substrate (schematic in Fig. 4a). The resonant peak shifts $\Delta\omega = 815\text{ cm}^{-1}$ when $\varepsilon$ changes from 1 to 3, as shown by the dashed arrow between the maxima of the green and black curves in Fig. 4b. For comparison, in Fig. 4d we show an analogous set of calculations for a GP resonance in a graphene disk lying on a transparent dielectric substrate ($\varepsilon = 1$) with the same thickness (2 nm) as the spacer in the AGP nanoresonator (schematic in Fig. 4c). According to Fig. 4d, the resonant peak ($A_1'$ mode) shows significantly smaller shift of $\Delta\omega = 168\text{ cm}^{-1}$ for $\varepsilon$ changing from 1 to 3 compared to AGP nanoresonator. For further quantitative characterization of the sensitivity of the AGP nanoresonator to the change of $\varepsilon$, we introduce a figure of merit (FOM), defined as $f = \frac{\Delta\lambda}{\text{FWHM}\cdot\Delta n}$, where FWHM is the full width at half maximum of the resonance peak (taken for either $A_1$ or $A_1'$ modes), while $\Delta\lambda$ is the wavelength shift of resonant spectral position and $\Delta n$ is refractive index change ($n = \sqrt{\varepsilon}$). The calculated FOM of the AGP disk nanoresonator ($f = 11.4$) is larger by a factor of 7 compared to the FOM of the GP disk nanoresonator ($f = 1.5$), thus demonstrating a much higher sensing ability of AGP nanoresonators compared to GP nanoresonators. Moreover, the FOM of AGP nanoresonantors is higher than that of the conventional sensing configurations in the mid-IR, e.g. the metallic rod antenna ($f = 9$)[38,39] or semiconductor grating ($f < 1$)[40] where the refractive index of the semi-infinite substrate or that of the thick film is probed. These results demonstrate that AGP nanoresonators possess high FOM corresponding to small volumes ($2 \cdot 25^2 \cdot \pi\ nm^3$ in case illustrated in Fig. 4a) of the probed material, revealing a clear advantage of AGP nanoresonator-based sensors over the conventional and GP nanoresonators based-mid-IR sensors.

**Enhanced molecular vibrational fingerprints with AGP nanoresonators**

Apart from refractive index sensing, the strong plasmonic field confinement inside the AGP nanoresonator can be exploited to perform enhanced IR absorption spectroscopy of vibrations of nanoscale volumes. Conventional measurements of extinction spectra are not efficient due to a weak coupling of AGP nanoresonators to the incident free-space radiation. Therefore, excitation of AGP nanoresonators by localized sources, e.g. metallic antenna,[41] would be a much better scenario for a practical realization of sensing. An example of mid-IR sensing with the help of metallic antenna is Fourier transform IR nanospectroscopy (nano-FTIR).[42] The nano-FTIR technique is based on a scattering-type scanning near-field optical microscope (s-SNOM) coupled to a Fourier spectrometer, where a metalized amplitude force microscopy tip (playing the role of an IR antenna) is illuminated by a broadband source. The light back-scattered from the tip (tip-scattered signal) carries information on the tip-sample near-field interaction taking place at the nanometric tip apex. Therefore, the tip-scattered signal can sense chemical composition of the sample with the nanoscale resolution given by the tip apex radius. For typical apex radii of the commercially available tips (r~30 − 50 nm), the momentum of AGPs belongs to the ample range of momenta k provided by these tips (limited to $k_{max} \sim 1/r$), enabling the coupling of the tip near fields with AGPs. In order to study the vibrational sensing capabilities of the AGP nanoresonators, we simulate a nano-FTIR-like experiment, assuming that the vertical antenna (metallic tip) excites the AGP nanoresonator. Tip-scattered signal can be qualitatively approximated by simulating vertical electric field $E_z$ below a dipole source placed above the sample.[24, 43] The testing material that we use to investigate AGP nanoresonator for sensing is a layer of proteins.[3] The testing molecule has two well-defined IR vibrational bands (amide I and II), as clearly visible in its dielectric function $\varepsilon_{pro}$, plotted in Fig. 5e. To evaluate the sensitivity, we compare three different sensing schemes, sketched in Fig. 5: small area ("protein disk" of diameter 53 nm) of 2 nm thick protein layer directly on top of the gold film (Fig. 5b), protein disk below a conventional GP nanoresonator of diameter 205 nm on a dielectric substrate (Fig. 5c) and a protein disk with diameter 53 nm in the spacer of AGP nanoresonantor (Fig. 5d). The resulting nano-FTIR spectra (see Methods) are plotted in Fig. 5f. The spectrum for the bare protein disk (black line) barely shows the amide I and II bands signature. The weakness of the protein signatures obviously makes it challenging to experimentally detect the protein disk since such a small-amplitude feature in the scattered signal can easily be below the noise level. The conventional GP nanoresonator does not significantly improve the visibility of the protein fingerprint. Indeed, the red line in Fig. 5f does not reveal any appreciable feature so that the spectrum resembles the one of GP nanoresonator without the protein disk.[24] This is confirmed by the Fig. 5f, where the shape of the simulated nano-FTIR spectrum for the GP nanoresonator with the protein disk (red curve) is similar to the one for the AGP with the disk of a constant dielectric permittivity $\varepsilon_{ref}$ (shown in Fig. 5e by the dashed red line), in which the vibrational modes have been artificially removed (purple dashed curve in Fig. 5f). The situation completely changes for the case of the protein disk inside the AGP nanoresonantor (blue curve in Fig. 5f). The nano-FTIR spectrum is dramatically affected by the amide bands, whose fingerprints are straightforwardly recognizable in the AGP spectrum. Comparison with the reference spectrum calculated with $\varepsilon_{ref}$ makes the appearance of the protein fingerprints more evident. Such a remarkable vibrational signature for an extremely small molecular volume ($2 \cdot 26.5^2 \cdot \pi \, nm^3$) makes AGP resonators a very promising platform for enhanced IR absorption spectroscopy applications.

It is important to highlight that in AGP nanoresonantors, the mode volume is approximately equal to the volume of the spacer in AGP nanoresonator and can have very small value. For example, in the AGP nanoresonator shown in Fig. 5, the mode volume is roughly $10^{-8}$ times smaller than the cube of the wavelength in free space, $\lambda_0^3$. Such extremely small mode volume can be promising for achieving the strong coupling of AGPs with molecular vibrations. In order to achieve the strong coupling, the AGP resonance width should match with that of the molecule absorption line. We estimate the absorption line width of molecule shown in Fig. 5a (analyzing the peaks in $\text{Im}(\varepsilon_{\text{pro}})$) to be 105 cm$^{-1}$. The same AGP resonance width corresponds to the relaxation time of the charge carriers in graphene of $\tau = 0.1$ ps (or mobility of 2000 cm$^2$/(V•s) for $E_F = 0.5$ eV).[24,30-32] Thus, the observation strong coupling between the AGP nanoresonators and molecular vibrations can be realistic even for standard graphene samples of low-mobility. Note that the quality of both metallic substrate and graphene sheet have still to be high enough (in order to avoid heterogeneity effects), as well as in surface enhanced infrared absorption spectroscopy.[44-46] For example, usage of single crystalline gold for the samples[47] would be preferable.

**CONCLUSION**

In conclusion, we have investigated the acoustic graphene plasmons in graphene disk nanoresonator placed above a metallic substrate. We have found that such a resonator can support many electromagnetic modes which can be categorized into the edge and sheet acoustic plasmons. Due to the ultra-confined electric field of all the acoustic modes, the spontaneous decay rate of a point emitter placed close to the graphene disk is enhanced up to 8 orders of magnitude compared to the decay rate in free space. We suggest the AGP resonator to perform as a sensor of ultra-small volumes of analytes, in particular small quantities of molecules. We show that the sensitivity of AGP nanoresonators is much higher than that of the conventional GP nanoresonators, with a figure of merit larger by a factor of 7. We further demonstrate that this nanoresonator can be used for enhanced IR vibration spectroscopy by simulating spectroscopic signatures of a protein placed in the AGP resonator gap. We believe that our findings are of great importance for the IR sensing, particularly for the ultra-small volume of molecules, and enhanced light-matter interactions, such as strong coupling.

**METHODS**

**Full wave simulations.** The calculations have been performed with the help of finite element methods using Comsol software. In simulations, the scattering boundary condition has been used and the graphene layer has been modeled as a surface current in the boundary conditions. The graphene is treated as a two-dimensional layer with the frequency-dependent conductivity $\delta(\omega)$, which we take according to local random phase approximation.[28, 48, 49] The wavelengths-dependent dielectric function of gold is taken from Ref. 50.[50] The Fermi energy of the doped graphene is $E_F = 0.5$ eV and relaxation time is $\tau = 0.05$ ps.

**The decay rate**. According to Poynting's theorem the radiated power of a current source is identical to the rate of energy dissipation (the decay rate), $dW/dt$. The latter is given by the following equation,[51] $\Gamma = dW/dt = -\frac{1}{2}\int dV \, \text{Re}\left(\frac{4\pi}{c} j^* \cdot E\right)$, where j and E are the current distribution and the electric field, respectively, and the integral is taken over the whole space (note that in contrast to Ref. 51, the factor $\frac{4\pi}{c}$ appears due to the Gauss unit system use). Assuming that the current distribution of a point source (located at $r = r_0$) with the dipole moment parallel to z-axis is $j = e_z\delta(r - r_0)$, the decay rate becomes $\Gamma = -\frac{2\pi}{c}\text{Re}[E_z(r_0)]$. The decay rate of the same point source in free space, $\Gamma_0$, is given by $\Gamma_0 = 4\pi\omega^2/3c^4$, so that dividing $\Gamma$ by $\Gamma_0$, we arrive at the Purcell's factor given by Eq.1. In the literature, the energy loss of a point source is commonly expressed via the Green's dyadic $\widehat{G}(r, r')$. This relation can be derived from the Poynting's theorem as well, remembering that the field generated by a current distribution is given by $E(r) = \frac{4\pi i\omega}{c^2}\int dV' \widehat{G}(r, r')j(r)$. Substituting here the above current distribution for the point source and then plugging $E(r)$ into the expression for the energy dissipation, the latter becomes $\Gamma = \frac{4\pi i\omega}{c^2}\text{Im}[G_{zz}(r_0, r_0)]$.

**Protein permittivity model.** The permittivity of the real protein used in this work is set as $\varepsilon(\omega) = n_\infty^2 + \sum_{k=1}^{2}\frac{s_k^2}{\omega_k^2 - \omega^2 - i\omega\gamma_k}$, and the parameters in the protein permittivity are: $n_\infty^2 = 2.08$, $\omega_1 = 1668 \text{ cm}^{-1}$, $\omega_2 = 1532 \text{ cm}^{-1}$, $\gamma_1 = 78.1 \text{ cm}^{-1}$, $\gamma_2 = 101 \text{ cm}^{-1}$, $S_1 = 213 \text{ cm}^{-1}$ and $S_2 = 200 \text{ cm}^{-1}$.

**Modeling of s-SNOM with a point dipole.** In Fig. 5, the electric-field enhancement is defined as $|E_z/E_{z,\text{sub}}|$, where $E_{z,\text{sub}}$ refers to the electric field of the gold film and $E_z$ refers to the electric field for GP or AGP nanoresontors. Here, the spectra for AGP and GP nanoresonator have respectively been normalized by using the maximum field enhanced value, which is generated under the case of the reference layer ($\varepsilon = \varepsilon_{\text{ref}}$). The spectrum for the protein-gold film is normalized by its maximum enhanced value. Here the electric dipole locates at the center of the graphene disk simultaneously with 50 nm of z-position. The detected point of electric-field enhancement locates at $z = 5$ nm, simultaneously keeping the same x- and y-positions as the dipole. It is important to highlight that, the reconstruction of the dielectric function of the molecular layer can only be done by taking into account the phase of the near fields. A complex interference between the near fields of the tip, excited plasmonic AGP mode and molecular vibrations leads to the Fano profile[24] (see blue solid curve in Fig. 5f). Therefore, a straightforward subtraction of the signal amplitudes for the spectra with and without molecules will not give the dielectric function.

# AUTHOR INFORMATION

**Corresponding Author**
*E-mail: a.nikitin@nanogune.eu and r.hillenbrand@nanogune.eu

**Notes**
The authors declare no competing financial interest.

# ACKNOWLEDGMENTS

The authors acknowledge support from the European Commission under the Graphene Flagship

(GrapheneCore1, grant no. 696656), and the Spanish Ministry of Economy and Competitiveness (national projects MAT2014-53432-C5-4-R and MAT2015-65525- R). S.C. and Z. L. Y. acknowledge the National Natural Science Foundation of China (No.11474239, No.21673192) and MOST of China (No.2016YFA0200601). We thank Dr. M. Okuda for help with the illustrations.

33. Notice that in the plasmonic region (ii), where the dipole transfers all its electromagnetic energy to plasmons, the γ reaches its maximal value of $6\times10^6$, so that the distance z = 5 nm used in Fig. 3

**Table of Contents**

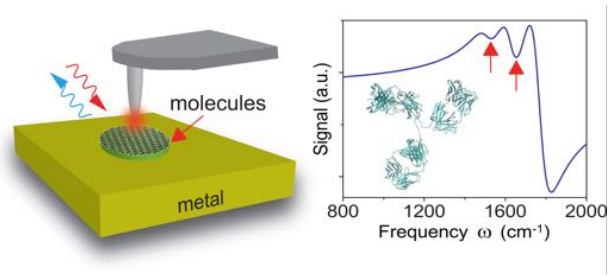

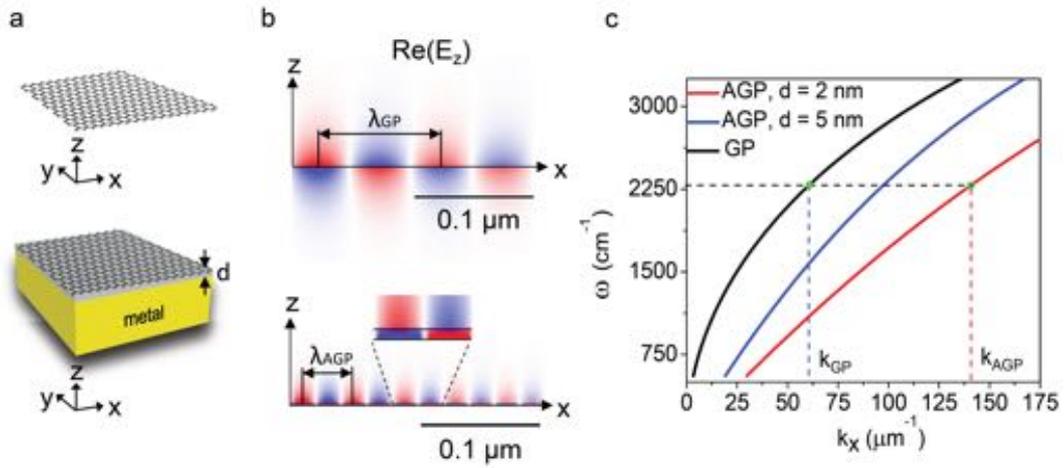

**Figure 1**. Field distribution and dispersion for GPs and AGPs. (a) Schematic of infinite graphene sheet (upper panel) and infinite graphene sheet above metal (bottom panel). (b) Spatial distribution of $\mathrm{Re}(E_z)$ for GPs (upper panel) and AGPs (bottom panel) at the frequency $\omega = 2292 \text{ cm}^{-1}$. The distance of the air spacer is $d = 2 \text{ nm}$. The horizontal black arrows indicate the wavelength of both, GPs, $\lambda_{GP}$, and AGPs, $\lambda_{AGP}$. The inset shows the enlarged field distribution in the 2 nm gap region. (c) Dispersion relation of GPs (black curve) and AGPs for $d = 2 \text{ nm}$ (red curve) and $d = 5 \text{ nm}$ (blue curve). The dotted lines mark the values of $k_{GP}$ and $k_{AGP}$ at the fixed frequency $\omega = 2292 \text{ cm}^{-1}$.

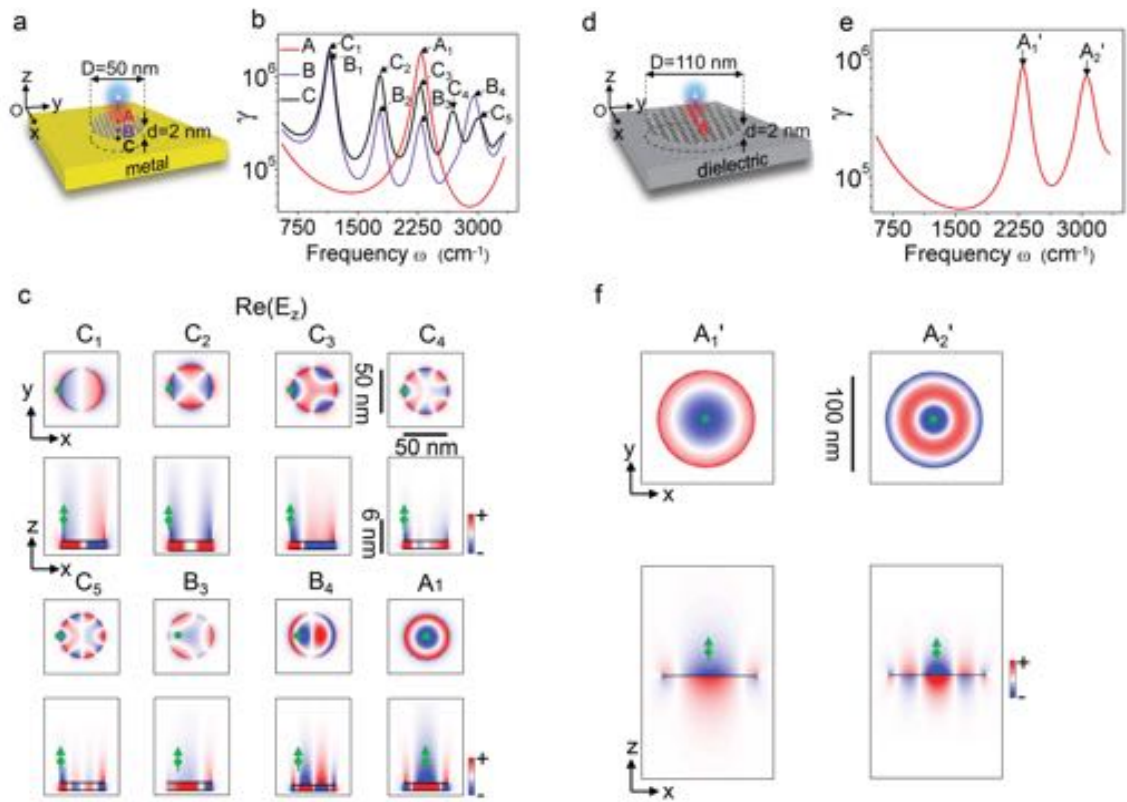

**Figure 2.** Coupling of an emitter to AGP and GP disk nanoresonators. (a) Schematics of the AGP disk nanoresonator with 2 nm air spacer. The size of the nanodisk is 50 nm. The capitals A, B and C label the location of dipole at center (A), between the center and edge (B) and the edge (C), respectively. (b). The normalized total decay rate of the emitter for three different locations, A (red curve), B (purple curve) and C (black curve). (c) Spatial distribution of $Re(E_z)$ at the xy-plane (top) and xz-plane (bottom) of different modes corresponding to the resonant peaks in (b). (d) Schematics of graphene nanodisk lying on a dielectric substrate (we take the refractive index to be 1). The size of the nanodisk is 110 nm. (e) The normalized total decay rate of the emitter at the center of the nanodisk on the dielectric substrate. (f) Spatial distribution of $Re(E_z)$ at the xy-plane (top) and xz-plane (bottom) for different modes corresponding to the resonant peaks in (e). The dipole source (located 5 nm above the nanodisk) is represented by the green arrow with a dot.

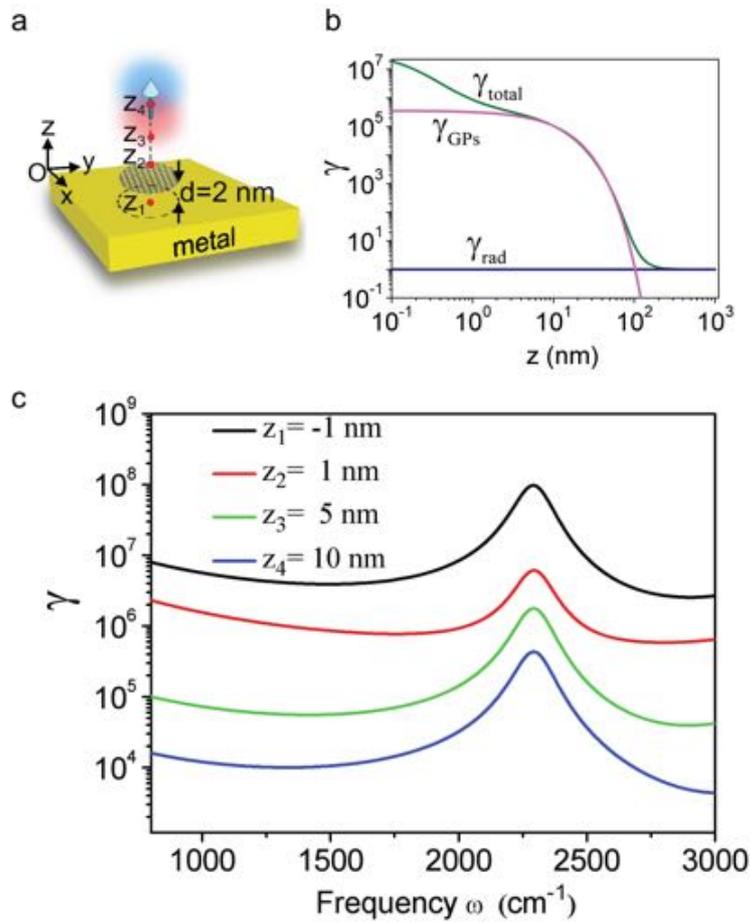

**Figure 3.** The normalized total decay rate of an emitter as a function of the distance z between the emitter and graphene disk. (a) Schematic of the doped graphene nanodisk above a metallic substrate with a point dipole source located at different distance ($z_1$, $z_2$, etc.) above the disk. The diameter of the graphene nanodisk is 50 nm, and the thickness of the air spacer between the metal and the disk is 2 nm. (b) The dependence of the normalized decay rate on z at the resonant frequency, $\omega = 2292$ cm$^{-1}$. (c) The normalized total decay rate of the emitter at different (z): $z_1 = -1$ nm (black), $z_2 = 1$ nm (red), $z_3 = 5$ nm (green) and $z_4 = 10$ nm (blue), as the schematic shown in (a).

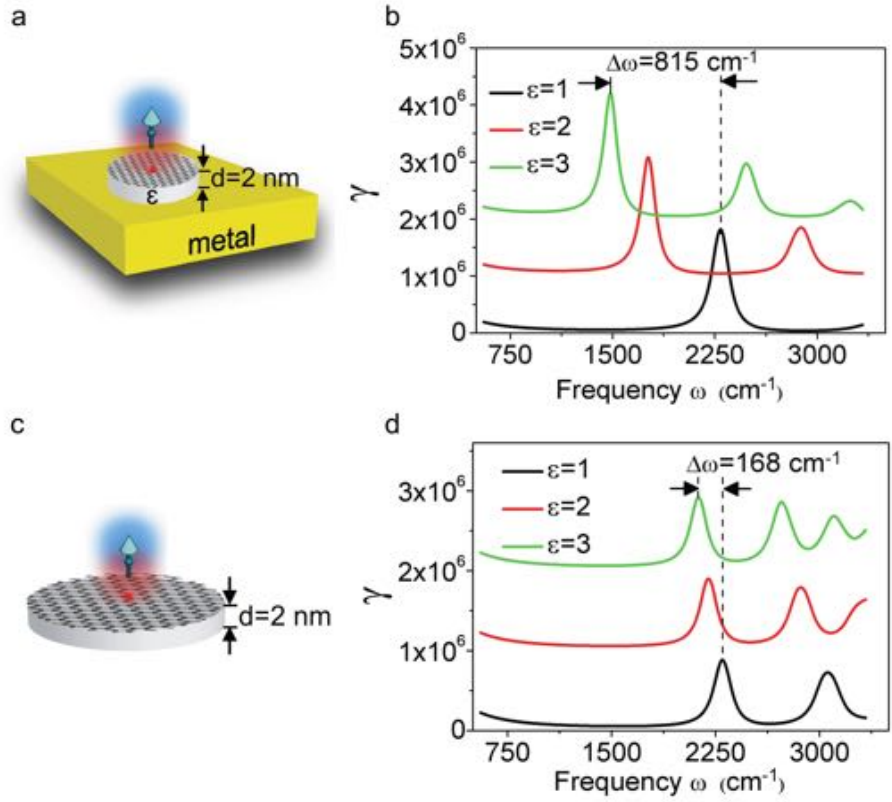

**Figure 4.** The dependence of the normalized total decay rate of AGP and GP disk nanoresonators upon the dielectric permittivity, $\varepsilon$. (a) and (c) Schematics for AGP disk nanoresonator with 50 nm diameter and GP disk nanoresonator with 110 nm diameter. The thickness of the dielectric disk is 2 nm in both configurations. (b) and (d) The normalized total decay rate for the configuration (a) and (c), respectively. $\varepsilon = 1$: black curve, $\varepsilon = 2$: red curve, and $\varepsilon = 3$: green curve.

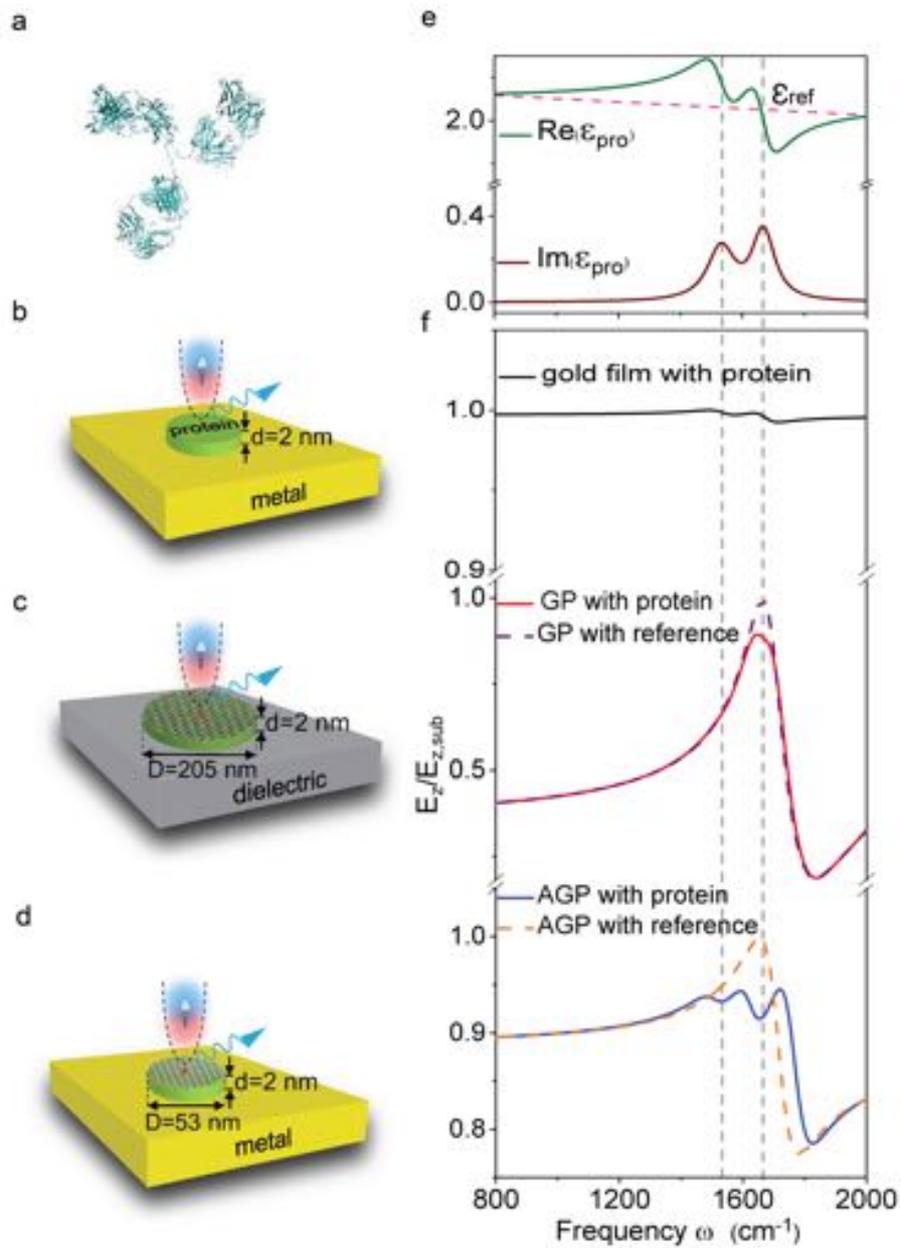

**Figure 5.** AGP disk nanoresonator for IR sensing of protein molecule fingerprint. The schematics of: (a) protein molecule, (b) protein molecules disk with 53 nm diameter on the gold film, (c) graphene disk resonator with 205 nm diameter-protein molecules-dielectric film system and (d) AGP nanoresonantor with 53 nm diameter in which the gap is inserted by the protein molecules. The real part of the dielectric permittivity of the protein, Re($\varepsilon_{pro}$) (green curve), and the imaginary part, Im($\varepsilon_{pro}$) (brown curve) as a function of frequency. Pink line shows the dielectric permittivity of the reference layer, $\varepsilon_{ref}$. (f) The normalized electric-field spectra for the configurations shown in: b (black curve), c (red curve) and d (blue curve) respectively. The purple and orange dotted curves represent the normalized electric-field spectra of the configurations in c and d with reference layer, respectively. The two grey dotted lines indicate the spectral positions of the two vibrational absorption peaks of the protein molecule. The dipole approximating s-SNOM tip is located at 50 nm above the studied configurations, while the near-field is calculated 5 nm above the graphene.